\documentclass{aa}  
\usepackage{graphicx}
\usepackage{xcolor}
\usepackage{caption}
\usepackage{subcaption}
\usepackage{makecell}
\graphicspath{ {./images/} }
\usepackage{nccmath}
\usepackage{txfonts}
\begin{document}

   \title{Asteroid pairs: method validation and new candidates}

   \author{I. Kyrylenko
          \inst{1}
          \and
          Yu. N. Krugly\inst{1}
          \and
          O. Golubov\inst{1}
          }

   \institute{Institute of Astronomy, 
              V. N. Karazin Kharkiv National University,
              35 Sumska Str.,  
              Kharkiv,
              61022,
              Ukraine\\
              \email{ihor.kyrylenko@karazin.ua}
             }

   \date{Received January 20, 2021; accepted July 19, 2021}

  \abstract
   {An asteroid pair can be described as two asteroids with highly similar heliocentric orbits that are genetically related but not gravitationally bound. They can be produced by asteroid collisions or rotational fission. Although over 200 asteroid pairs are known, many more are remaining to be identified, especially among the newly discovered asteroids.}
   {The purpose of our work is to find new asteroid pairs in the inner part of the main belt with a new pipeline for asteroid pair search, as well as to validate the pipeline on a sample of known asteroid pairs.}
   {Initially, we select pair candidates in the five-dimensional space of osculating orbital elements. Then the candidates are confirmed using numerical modeling with the backtrack integration of their orbits including the perturbations from the largest main-belt asteroids, as well as the influence of the non-gravitational Yarkovsky effect.}
   {We performed a survey of the inner part of the main belt and found 10 new probable asteroid pairs. Their estimated formation ages lie between 30 and 400 kyr. In addition, our pipeline was tested on a sample of 17 known pairs, and our age estimates agreed with the ones indicated in literature in most of the cases.}
   {}


   \maketitle
%

\section{Introduction}

    In the past few decades, a lot of pairs have been found among the main belt asteroids, members of which possess highly similar orbits and share a common origin. 
    
    There are two major ways of how a pair is created. First, a collision of two asteroids can shatter them into fragments \cite{Michel2015}. Second, a rapidly rotating asteroid can undergo the rotational fission \citep{Scheeres2007, Vokrouhlicky2015} due to the centrifugal forces. The rapid rotation can be caused by the YORP effect, which is the radiation pressure torque acting on an asteroid of an asymmetric shape. One possible outcome of such rotational fission is a decay of an asteroid into several gravitationally unbound fragments \citep{Jacobson2011}. Both collisional disruption and rotational fission can create clusters with multiple members of different sizes \citep{Pravec_cluster2018}. In this article, we do not look for such clusters, but merely for pairs. This leaves open the possibility that some of the pairs we find could be in fact some of the largest members of certain clusters. 
    
    The two members of a newly formed asteroid pair initially have a small relative distance and velocity. Over time, their orbits drift apart due to perturbations from planets, and to a lesser extent due to the non-gravitational Yarkovsky effect. Even if the components of the pair are no longer bound gravitationally, their relation can be discovered via various methods. Thus the task emerges to identify such a pair and to uncover its dynamic history.
    
    One of the first successful attempts to discover asteroid pairs was made in \citet{Vokrouh2008}, where pair candidates were pre-selected based on their proximity in the five-dimensional space of osculating orbital elements, and then confirmed by the orbital integration. After that, a number of dynamical studies were published \citep{Pravec2009, Pairs2019} that have massively expanded the list of identified pairs.
    Other studies have shown the necessity of taking into account the most massive perturbers of the main belt (Ceres, Vesta, Pallas) in numerical calculations of the formation age of pairs \citep{Galad2012}. At the same time, the findings by \citep{Vokrouh2008} and \citep{Zizka2016} opened the road to the identification of very young asteroid pairs. Recent advancements in numerical propagation software \citep{Rebound_general} -- the main instrument in asteroid pairs identification -- made it easier to contribute to asteroid pairs search.
    
    We began our work in asteroid pair search with the development of a comprehensive software pipeline, both for the identification and verification of pair candidates. We have chosen the inner part of the main belt (2.0-2.5 a.u.) as the area of our search. The density of discovered asteroids is the highest in this region, which increases the chances of finding an asteroid pair. Moreover, the high collision rates and short YORP timescales guarantee a high pair production rate.
    
    We present the first results of our survey in this paper. Section 2 describes methods of asteroid pairs search and verification that were used in this article. Section 3 covers the results, namely 10 newly found probable asteroid pairs, whereas Section 4 describes the outcome of this work in general, alongside our further plans.

\section{Research methodology}

   \begin{figure*}
   \centering
   \includegraphics[width=18cm]{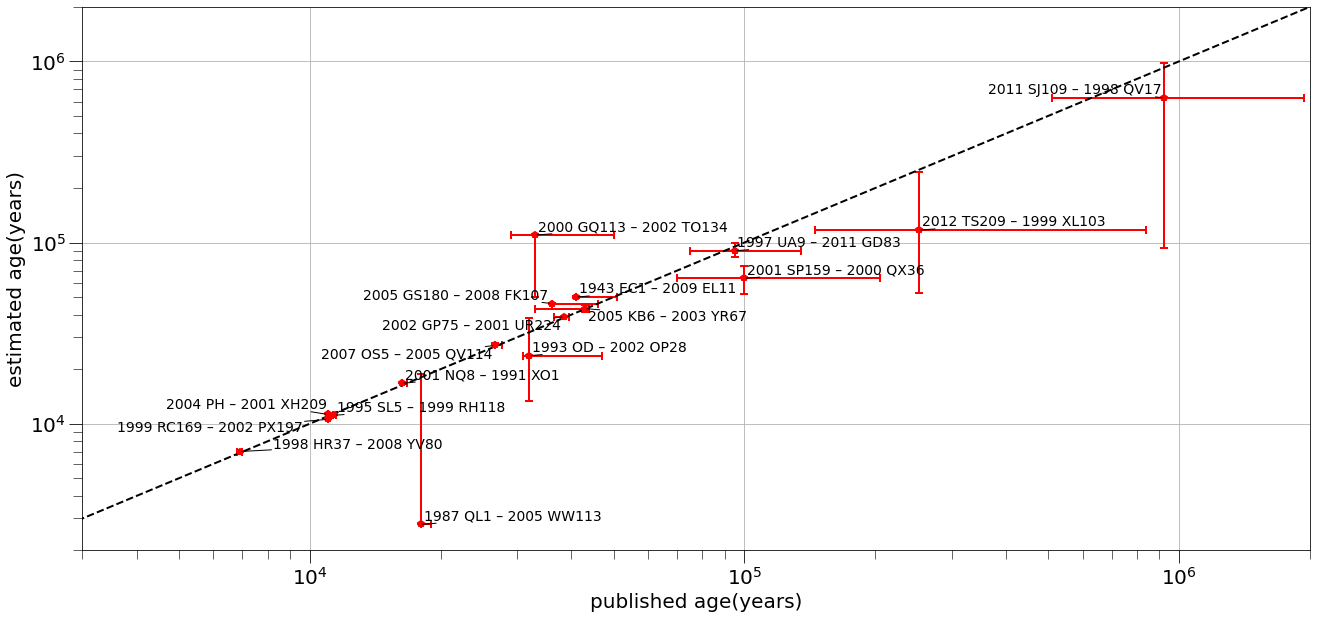}
        \caption{\begin{footnotesize} Age comparison between our pipeline and the literature. The $x$-axis of the plot shows the formation times of these pairs taken from the literature, and the $y$-axis shows the corresponding times calculated by our program. Red lines mark errorbars for both cases. The black dashed line represents the case when both results are in perfect agreement.\end{footnotesize}}
        \label{Known_pairs}
    \end{figure*}
\subsection{Initial pair search in five-dimensional space}
   To identify asteroids as a potential pair, we use a modified five-dimensional version \citep{Vokrouh2008} of the three-dimensional metric by \cite{zappala1990}, who applied it to asteroid families via HCM (hierarchical clustering method). In contrast to the search for asteroid families, where HCM is used with proper orbital elements of the asteroids, we use osculating orbital elements for our search. The point is that typical asteroid families are old enough to have erased any initial grouping in the phase space of the osculating elements, whereas the pairs we are searching for are sufficiently young to still retain the close resemblance even in the osculating elements. The main advantage of using the osculating orbital elements is that the longitude of the ascending node and the argument of perihelion preserve additional information about the orbit similarity. This abates the number of false pair candidates, which would have inevitably appeared while using only proper elements of the asteroids. As it was shown in \citet{Nesvorn2006} this approach allows us to reliably recover pairs with formation times which are $\lesssim$1 million years. The method consists of calculating the distance $d$ between asteroids in the five-dimensional space of the Keplerian orbital elements using the following formula: 
   \begin{equation}
   \Big(\frac{d}{na}\Big)^2
   = k_\mathrm{a}\Big(\frac{\delta a}{a}\Big)^2
   + k_\mathrm{e}({\delta e})^2
   + k_\mathrm{i}({\delta \sin{i}})^2
   + k_\mathrm{\Omega}({\delta \Omega})^2
   + k_\mathrm{\varpi}({\delta \varpi})^2.
   \label{five-dimensional distance}
   \end{equation}
   Here the semi-major axis $a$, the eccentricity $e$, the inclination $i$, the longitude of the ascending node $\Omega$, and the longitude of perihelion $\varpi$ are the Keplerian elements, ($\delta a$, $\delta e$, $\delta \sin{i}$, $\delta\varpi$, $\delta\Omega$) is the separation vector of neighboring bodies in the phase space, $n$ is the mean motion of the asteroid. Following \cite{Nesvorn2006}, we adopted the next values of the coefficients for our computations: $k_\mathrm{a} = 5/4$, $k_\mathrm{e} = k_\mathrm{i} = 2$, and for $k_\mathrm{\Omega} = k_\mathrm{\bar\omega} = 10^{-4}$. 
   
   The smaller the phase distance between given asteroids, the more their orbits are similar, and it is more likely that such a combination of asteroids may turn out to be evolutionarily related and gravitationally bound in the past. In our search, we limited the cutoff distance in the phase-space to $d=100$ m/s. On the one hand, for larger phase distances young asteroid pairs are less probable, on the other one, it is larger than the value that is commonly used \citep{Pairs2019}, giving the opportunity to find new pair candidates in well-studied regions.
   
   For our calculations, we used the osculating orbital elements provided by the JPL Horizons system\footnote{JPL Small-Body Database,\\\texttt{https://ssd.jpl.nasa.gov/sbdb.cgi}}. The values of orbital elements are given for the same epoch. The phase distances for potential pairs were also calculated for several MPC databases\footnote{Minor Planet Center, \\ \texttt{https://minorplanetcenter.net/iau/mpc.html}} covering years 2004-2019, and then the median value of the phase distance was calculated. This potentially makes our calculation less susceptible to errors of individual databases and short-term perturbations. In total, 15 MPC databases were used, which are mostly equidistant in time. As the result, $\sim500$ potential pairs were found with the median distance of up to 100 m/s in the phase space.

   Since the pair candidates found also contain confirmed pairs, we cross-match the obtained pair candidates with the known pairs in order not to waste the resources on their simulation. To do this, we use the Johnston database of known pairs \citep{Johnston_pairs} as well as small catalogs from \citet{Pairs2019}. Out of over 500 pair candidates found, 67 were identified as the known ones, which present more than 90 percent (67/72) of the known pairs listed in the Johnston database. Since the list of known pairs is constantly being updated, each studied pair is checked by NASA ADS\footnote{NASA astrophysics data system,\\ \texttt{https://ui.adsabs.harvard.edu/}} for related articles. Pair candidates that were not mentioned in any source were selected for further processing.

\subsection{Numerical modeling}

    For further verification of the obtained potential pairs, we use numerical modeling. The principle of verification is to use backtrack integration for the components of the pair to find their close approaches, which are characterized by the small relative velocities and distances. For numerical modeling, we use REBOUND package, which is available for Python and C programming languages \citep{Rebound_general}. REBOUND package provides a variety of N-body integrators, out of which we used WHFast, SABA, and IAS15. The WHFast and SABA integrators \citep{Rebound_WHFast} were applied primarily for the pairs with unambiguous and unique encounters, whereas IAS15 \citep{Rebound_IAS15} was used for pairs with more complex dynamical behavior. Although IAS15 is distinctly slower, its high-order nature with the adaptive integration timestep gives optimal results down to machine precision for errors associated with numerical integration. The length of the time step used in integrations is 1 to 3.65 days for WHFast and SABA integrators, whereas the IAS15 minimal time step was limited to 0.5 days. The time step was chosen for each pair individually, depending on the radius of the Hill sphere of the largest component of the pair, and the mutual velocity of the clones during the encounters.

    As a frame for computations, we build a model that consists of all the planets of the Solar System. The most massive bodies of the main-belt, which are Ceres, Vesta, and Pallas, were also taken into account. These bodies besides planets are the principal main belt perturbers, and can significantly influence the behavior of asteroids, thus modifying the determined age of the pair \citep{Galad2012}. Planets with moons are represented by their barycenters.
    
    Since the orbital elements of asteroids are known with some uncertainty, which is associated with the accuracy of observations, it is necessary to take into account these errors when adding asteroids to the simulation. To implement this, we generate an array of asteroid clones with the values of orbital elements within the given accuracy. The orbital elements, as well as their uncertainties, were taken from the JPL Horizons system. To consider the possible correlation between orbital parameters, we incorporated the mechanism for asteroid clones creation that was proposed in \citep{Rheinland2017}. The initial orbital elements of the clones $E$ are determined as
    \begin{ceqn}
        \begin{equation}
             E = T^{T}z + E_{*}, 
        \label{Covariation matrix}
        \end{equation}
    \end{ceqn}
    where $z$ is the six-dimensional vector whose components are random deviates of the normal distribution, and $E_{*}$ is the best-fit solution. The covariance matrix $T$ is obtained using the Cholesky decomposition method and satisfies $T^{T}T = \Sigma$, where $\Sigma$ is the normal matrix. JPL Horizons is not always able to provide covariance matrices and osculating orbital elements for the same epoch. To overcome this, clones of the asteroids are added to the simulation at their corresponding epochs as the integration into the past progresses.
  
    For an accurate calculation of the pair's evolution, especially in close approaches, it is important to take into account the mutual gravitational interaction of the pair's components. While simulating clones with masses makes it impossible to simulate more than two clones for a pair in one integration, it could give more robust results where multiple encounters occur. To do so, one must take into account the mass of asteroids. Since the mass for many asteroids is poorly constrained, in the first approximation the mass can be derived from the asteroid radius $R$ and bulk density $\rho$ using the formula: 
    \begin{ceqn}
        \begin{equation}
        m_\mathrm{ast}
        = 4/3 \pi R^3\rho,
        \label{asteroids mass}
        \end{equation}
    \end{ceqn}
    where the bulk density is assumed to be $2\cdot10^3\,\mathrm{kg\,m^{-3}}$ if not provided by the HORIZONS system. For asteroids whose radius is unknown, the radius is calculated using the asteroid's absolute magnitude $H$ and geometric albedo $p_v$ using the following formula \citep{Harris1997}:
    \begin{ceqn}
        \begin{equation}
        R_\mathrm{ast} = \frac{1329}{2\sqrt{p_v}}10^{-0.2H}  [\mathrm{km}]. 
        \label{asteroids radius}
        \end{equation}
    \end{ceqn}
    Mass is also required to calculate the Hill sphere radius and the escape velocity -- both of these parameters are used in the criteria for close approaches.
    
    The Hill sphere radius, which is the distance up to where the body dominates in the gravitational attraction, can be calculated using the equation:
    \begin{ceqn}
        \begin{equation}
        R_\mathrm{Hill} = a\sqrt[3]{\frac{{m}}{3M}}, 
        \label{Hill sphere radius}
        \end{equation}
    \end{ceqn}
    where $a$ and $m$ are the the semi-major axis and the mass of the primary component of the pair, and $M$ is the mass of the Sun.
    
    The escape velocity is the minimum speed needed to escape from the gravitational influence of a given body. Given the mass $m$ of the primary component and the distance $r$ between the center of mass of the primary and the secondary, the escape velocity is calculated as:
    \begin{ceqn}
        \begin{equation}
        v_{e} = \sqrt{\frac{{2Gm}}{r}}.
        \label{Escape velocity}
        \end{equation}
    \end{ceqn}
    
    We are mainly interested in those simulations in which close encounters between the components of a pair occur. We define such encounters by the limits of 5$R_\mathrm{Hill}$ and 2$v_{e}$ for the component with the largest mass \citep{Pairs2019}. For pair candidates with worse convergence, these conditions are extended up to 10-20$R_\mathrm{Hill}$ and 4-8 $v_{e}$ \citep{Pairs2019}. 

    The simulation also needs to take into account the Yarkovsky effect, which is a non-gravitational force arising due to non-isotropic thermal radiation from a rotating asteroid heated by the Sun. The result of the Yarkovsky effect can lead to the drift of the semi-major axis of the asteroid, either towards its increase for prograde rotators, or decrease in the case of retrograde rotators. This effect can strongly modify the behavior of an asteroid pair and make changes to its determined age. Since the physical and orbital parameters of an asteroid are often not very precise, the simplified formula for the Yarkovsky effect is sufficient. Using Eqn. (41) from \cite{golubov16} and integrating the pressure over the surface of an approximately spherical asteroid, we get the following expression for the Yarkovsky force (see Appendix \ref{AppendixYarkovsky} for derivation):
    
    \begin{ceqn}
        \begin{equation}
        F_\mathrm{Yark} = 0.28\frac{(1-A)\Phi R^2}{c}
        \label{yarkovs}
        \end{equation}
     \end{ceqn}
    Here $R$ is the radius of the asteroid, and $c$ is the speed of light. $\Phi=\frac{\Phi_0 a_0^2}{a^2}$ is the solar constant at the asteroid's orbit, $\Phi_0$ is the solar constant at $a_0=1$ AU, and $a$ is the semi-major axis of the asteroid's orbit. 
    
    To determine the direction of this force, it is also necessary to know the direction of rotation of the asteroid around its axis, which is also mostly unknown. For this reason, for each clone, we generate a random value of the Yarkovsky force which amounts to Eqn. (\ref{yarkovs}) multiplied by constant with uniform distribution in the range $[-1, 1]$. The YORP effect is not taken into account in this work, since we are looking for pairs of age $\leq$1M years, which is much less than the typical YORP timescales \citep{SheeresGolubov}.
    
    Although we are aiming at finding asteroid pairs with ages up to 1 million years, many of the pairs under consideration are much younger. Thus, to save the computation time, the reasonable choice is to limit the computational time to the latest possible encounter individually for every pair. In order to do it, we are checking relative distances and velocities for a sample of one thousand clones to determine time limits that would cover all the possible encounter times within 1 Myr interval. Then, having found these moments, we integrated 10-30 thousand clones in the obtained time intervals, including more clones for pairs with a worse convergence to compensate for larger dispersion.
    
    To check the efficacy and accuracy of our pipeline we compared its output with the results published in the literature. We took a sample of pairs with known formation ages, which were re-discovered in our survey, and recalculated their formation times. The comparison of our results and the results of other authors are shown in Figure \ref{Known_pairs} and Table \ref{Known_pairs_comparison}. Quite a good agreement of our result with the previous studies corroborates that our pipeline can be used to reliably discover pairs and to obtain their ages. The results are mostly within the known limits, although the errors can deviate from the ones cited in the literature for pairs with ambiguous multiple encounters.
    
    \begin{table*}[h]
    \captionsetup{justification=centering}
    \caption{Comparison of the determined formation ages between our pipeline and literature, previously shown in Fig.1. Data on pairs ages were taken from \cite{Johnston_pairs} and references therein.}
    \setlength{\tabcolsep}{14pt}
    \renewcommand{\arraystretch}{1.4}
    \centering
        \begin{tabular}{ |p{0.3cm}|p{2.5cm}|p{2.5cm}|p{2cm}|p{2cm}| }
         \hline
         \multicolumn{5}{|c|}{Known pairs age calculation} \\
         \hline
         No. & asteroid name & asteroid name & age literature [kyr] & estimated age [kyr]\\
         \hline
         1 & 1998 HR37 & 2008 YV80 & $6.9^{+0.1}$ & $7.0_{-0.1}^{+0.1}$ \\
         2 & 1999 RC169 & 2002 PX197 & $11_{-0.1}^{+0.1}$ & $10.8_{-0.1}^{+0.1}$\\
         3 & 2004 PH & 2001 XH209 & $11.1$ & $11.2_{-0.1}^{+0.3}$ \\
         4 & 1995 SL5 & 1999 RH118 & $11.4^{+0.1}_{-0.1}$ & $11.1^{+0.1}_{-0.1}$ \\
         5 & 1991 XO1 & 2001 NQ8 & $16.34^{+0.04}$ & $16.3^{+0.3}_{-0.5}$\\
         6 & 1987 QL1 & 2005 WW113 & $18.0^{+1}$ & $18.0^{+1}_{0.3}$ \\
         7 & 2007 OS5 &	2005 QV114 & $26.7^{+0.8}_{-0.2}$ & $27.0^{+1}_{-0.1}$ \\
         8 & 1993 OD & 2002 OP28 & $32.0^{+15.0}_{-1.0}$ & $23.7^{+14.5}_{-16.3}$ \\
         9 & 2000 GQ113 & 2002 TO134 & $33.0^{+17.0}_{-4.0}$ & $114.1^{+0.1}_{-0.1}$ \\
         10 & 2005 GS180 & 2008 FK107 &	$36.0^{+10}$ & $46.1^{+0.3}_{-0.2}$	\\
         11 & 2002 GP75 & 2001 UR224 & $40.0^{+10.0}$ & $38.9^{+0.2}_{-0.1}$ \\
         12 & 1943 EC1 & 2009 EL11 & $43.0^{+0.5}$ & $49.9^{+0.9}_{-0.7}$ \\
         13 & 2003 YR67 & 2005 KB6 & $43.0^{+1}_{-1}$ & $43.2^{+2}_{-1.7}$ \\
         14 & 1997 UA9 & 2011 GD83 & $95.0^{+40.0}_{-20.0}$ & $90^{+9.0}_{-6.0}$ \\	
         15 & 2001 SP159 & 2000 QX36 & $100.0^{+105}_{-30}$ & $63.5^{+0.1}_{-0.1}$ \\
         16 & 2012 TS209 & 1999 XL103 &	$252.0^{+586}_{-107}$ & $117.3^{+128.3}_{-64.2}$ \\
         17 & 2011 SJ109 & 1998 QV17 & $925^{+1014}_{-416}$ & $627.6^{+354.7}_{533.9}$ \\
         \hline
        \end{tabular}
    \label{Known_pairs_comparison}
    \end{table*}

\section{Results}
    
    Taking into consideration that the pipeline works correctly, we continued our search and found a number of pairs that satisfy the condition of close encounters during numerous backtrack integrations. 
    
    Following \cite{Pravec2009}, we checked the background around the obtained pairs for uniformity of its distribution and calculated the statistical significance of the pairs. We adopted the threshold values for the probability of a number of orbits surrounding the specific pair $P_{1/2}$, to be 0.01 for pairs with $d$ < 10 m/s, and 0.05 for pairs with greater $d$. Pairs with greater $P_{1/2}$ pass the test on the uniform distribution. As for the ratio, the probability of orbital coincidence of two genetically unrelated asteroids  $P_{2}/N_{p}$, pair is considered to be statistically significant if $P_{2}/N_{p} \ll 1 $. Some contamination by coincidental pairs is expected with $P_{2}/N_{p} = 0.1$ and larger.
    
    Orbital elements for the selected pairs are presented in Table \ref{Orbital_elements}, and results on estimated age formation together with significance estimates are in Table \ref{New_pairs_age}. 
    
    \begin{table*}[h]
    \captionsetup{justification=centering}
    \caption{Osculating orbital elements of the pair candidates for epoch MJD 59000.5. The bold digits indicate the current uncertainty of orbits, where the first uncertain digit is on the $1\sigma$ level.}
    \setlength{\tabcolsep}{15pt}
    \renewcommand{\arraystretch}{1.1}
    \centering
        \begin{tabular}{|@{\hskip0.2cm}p{0.01cm}|p{1.9cm}|@{\hskip0.3cm}p{1.5cm}|@{\hskip0.3cm}p{1.1cm}@{\hskip0.8cm}|@{\hskip0.3cm}p{1.2cm}|@{\hskip0.3cm}p{1.2cm}|@{\hskip0.3cm}p{1.2cm}|@{\hskip0.3cm}p{1.2cm}@{\hskip0.8cm}|}
         \hline
         \multicolumn{8}{|c|}{Osculating orbital elements of new pair candidates} \\
         \hline
         No. & asteroid name & $a$ (AU) & $e$ & $i$ (deg) & $\Omega$ (deg)  &  $\omega$ (deg) & $M$ (deg) \\
         \hline
         1 & 2004 RF90 & 2.3953751\textbf{48} & 0.2436204\textbf{8} & 7.10671\textbf{3} & 189.7046\textbf{1} & 117.2403\textbf{1} & 106.3064\textbf{98}\\
         - & 2003 UT336 & 2.394413\textbf{557} & 0.24406\textbf{554} & 7.0991\textbf{56} & 189.872\textbf{95} & 116.72\textbf{672} & 217.39\textbf{4766}\\ 
         
         2 & 2000 HS9 & 2.18629637\textbf{5} & 0.0702436\textbf{0} & 2.29516\textbf{3} & 83.637\textbf{316} & 88.510\textbf{470} & 121.3500\textbf{33} \\ 
         - & 2015 DF67 & 2.185110\textbf{452} & 0.071005\textbf{08} & 2.29454\textbf{4} & 83.631\textbf{496} & 88.038\textbf{373} & 209.722\textbf{014}	\\ 
         
         3 & 2003 RV20 & 2.3690030\textbf{86} & 0.1958752\textbf{1} & 2.33182\textbf{4} & 83.067\textbf{437}& 292.136\textbf{41} & 194.1996\textbf{08} \\ 
         - & 2010 TH35 & 2.369215\textbf{852} & 0.196216\textbf{19} & 2.3315\textbf{15} & 83.066\textbf{477} & 292.027\textbf{64} & 234.475\textbf{725} 	\\ 
         
         4 & 1999 WM4 & 2.28832493\textbf{0} & 0.0927631\textbf{5} & 5.29617\textbf{5} & 287.1019\textbf{3} & 150.6236\textbf{0}	& 326.2504\textbf{57} \\ 
         - & 2017 QD23 & 2.2893286\textbf{87}& 0.0924711\textbf{0} & 5.29424\textbf{7} & 287.0629\textbf{0} & 149.780\textbf{97} & 187.0416\textbf{91}\\ 
         
         5 & 2006 BJ193 & 2.2169224\textbf{40} & 0.1441583\textbf{8} & 1.76629\textbf{4} & 258.745\textbf{85} & 345.668\textbf{29} & 35.4789\textbf{228}\\
         - & 2017 FE106 & 2.2165703\textbf{87} & 0.143170\textbf{15} & 1.76911\textbf{6} & 258.829\textbf{21} & 345.553\textbf{73} & 306.5857\textbf{87}\\
         
         6 & 2015 VP32 & 2.3189024\textbf{10} & 0.226036\textbf{50}& 3.25439\textbf{4} & 308.7588\textbf{6} & 121.180\textbf{77} & 87.770\textbf{6531} 	\\ 
         - & 2002 CR55 & 2.31885178\textbf{0} & 0.2260049\textbf{0} & 3.26257\textbf{4} & 307.7059\textbf{8} & 122.4573\textbf{4} & 125.0182\textbf{73}\\ 
         
         7 & 2015 XO12 & 1.9793130\textbf{83} & 0.1014695\textbf{7} & 22.2885\textbf{8} & 59.1620\textbf{31} & 278.2336\textbf{1}& 268.8725\textbf{81}\\ 
         - & 2001 WY4 & 1.9795683\textbf{82} & 0.1014898\textbf{7} & 22.2871\textbf{2} & 59.1873\textbf{10} & 278.0949\textbf{9}	& 300.4133\textbf{03}\\
         
         8 & 1981 VL & 2.32652283\textbf{7} & 0.1556248\textbf{3} & 3.75908\textbf{1} & 24.1204\textbf{12} & 4.7306\textbf{743} & 321.80821\textbf{2}\\ 
         - & 2013 CX44 & 2.3263784\textbf{87} & 0.1564025\textbf{9} & 3.74960\textbf{9} & 23.767\textbf{811} & 5.634\textbf{6437} & 120.5590\textbf{82}\\ 
         
         9 &  1999 XF200 & 2.4362088\textbf{01} & 0.1649535\textbf{0} & 6.21759\textbf{7} & 324.1204\textbf{2} & 10.5475\textbf{51} & 229.9199\textbf{42}\\
         - & 2008 EL40 & 2.4373988\textbf{15} & 0.1652512\textbf{7} & 6.19744\textbf{4} & 323.2632\textbf{3} & 12.3513\textbf{69} & 293.2047\textbf{51}\\ 
         
         10 & 2000 XH16 & 2.4286958\textbf{04}	& 0.1664400\textbf{3} & 12.1904\textbf{8} & 294.0500\textbf{0} & 5.4335\textbf{170} & 156.9526\textbf{32} \\ 
         - & 2002 TM148 & 2.4277561\textbf{66}	& 0.1660767\textbf{2} & 12.1691\textbf{7} & 293.8474\textbf{8} & 5.0111\textbf{241} & 279.7400\textbf{00} \\
         \hline
         \end{tabular}
 \label{Orbital_elements}
 \end{table*}
 
Using the new database of SDSS colors for moving objects \citep{Sergeyev}, we were able to obtain the color indices for eight asteroids, with four of them being members of two pairs, allowing estimating feasibility of pairs not only with the orbital, but also with the optical properties.
This can be done by color indices $a_{*}$ and $i-z$  \citep{Ivezic2001}, or by $a_{*}$ and albedo \citep{Slyusarev2020}, with $a_{*}$ defined as
  \begin{ceqn}
        \begin{equation}
        a_{*} = 0.89\cdot(g-r) + 0.45\cdot(r-i) - 0.57.
        \label{Spectral classification}
        \end{equation}
    \end{ceqn}
Here $g$, $r$, $i$ are the magnitudes in the corresponding SDSS filters. The more similar are the parameters $a_{*}$ and $(i-z)$ for the components of the pair, the more it is probable that they are of the same spectral type.

While no information was found on spectral classes or color indices for the remaining pair candidates, we can base our calculations on the assumption that the components of the pair tend to be of the same spectral type, thus of the same albedo (if not provided by JPL HORIZONS system). With the available information on absolute magnitudes, we can roughly estimate size ratios for the components of the pair. 

We use the residual velocity and coordinate differences to make a preliminary guess of how the pair was created. A good convergence of coordinates and velocities simultaneously might be a manifestation of a pair formation via the YORP fission, whereas a relatively large velocity difference can imply a collisional scenario.
In some cases, especially for older pairs, improved orbit determination with the robust Yarkovsky effect will be needed to obtain more reliable results. 

\begin{figure*}[h!]
    \begin{minipage}[b]{0.49\linewidth} 
        \centering
        \includegraphics[width=\textwidth]{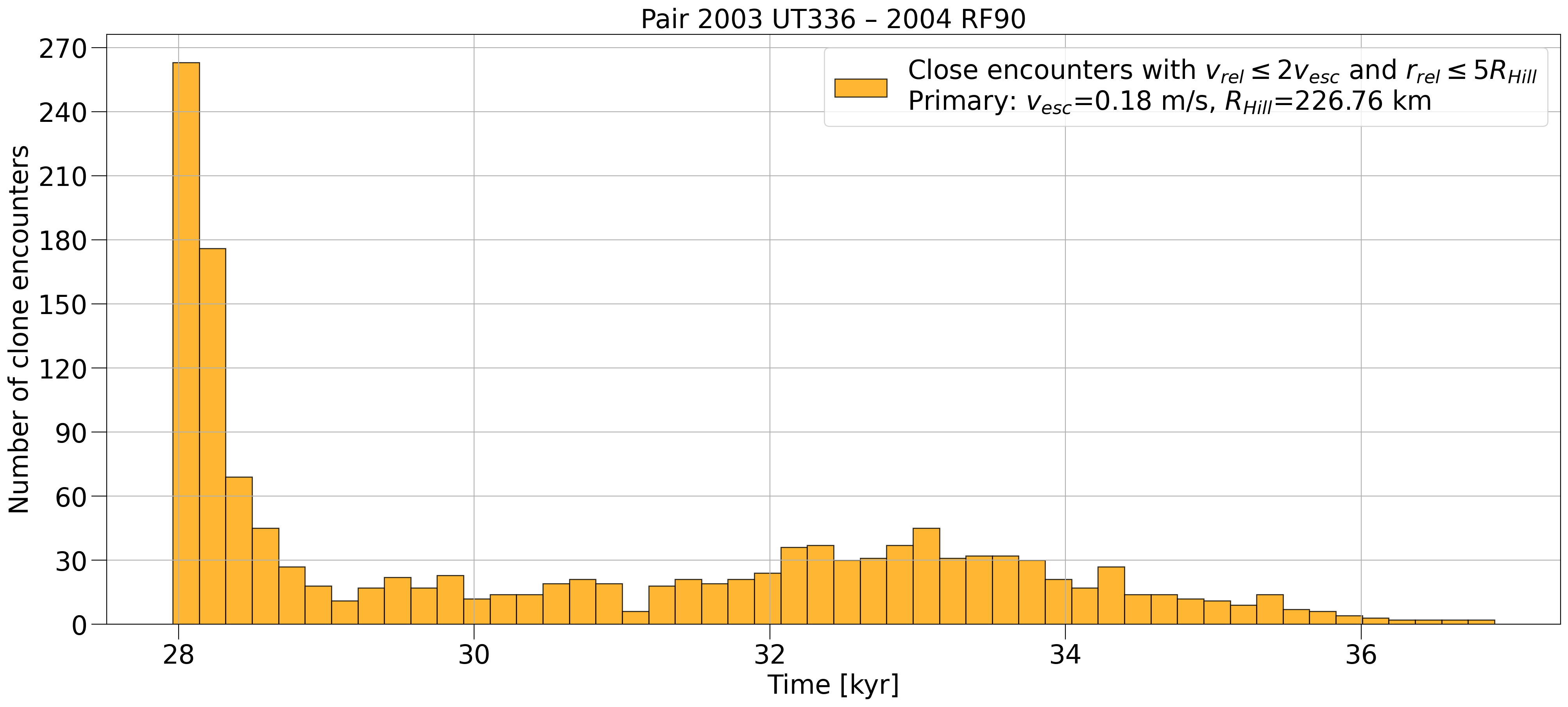}
        \label{fig:figure1}
        \end{minipage}
    \hspace{0.5cm}
     \begin{minipage}[b]{0.49\linewidth} 
        \centering
        \includegraphics[width=\textwidth]{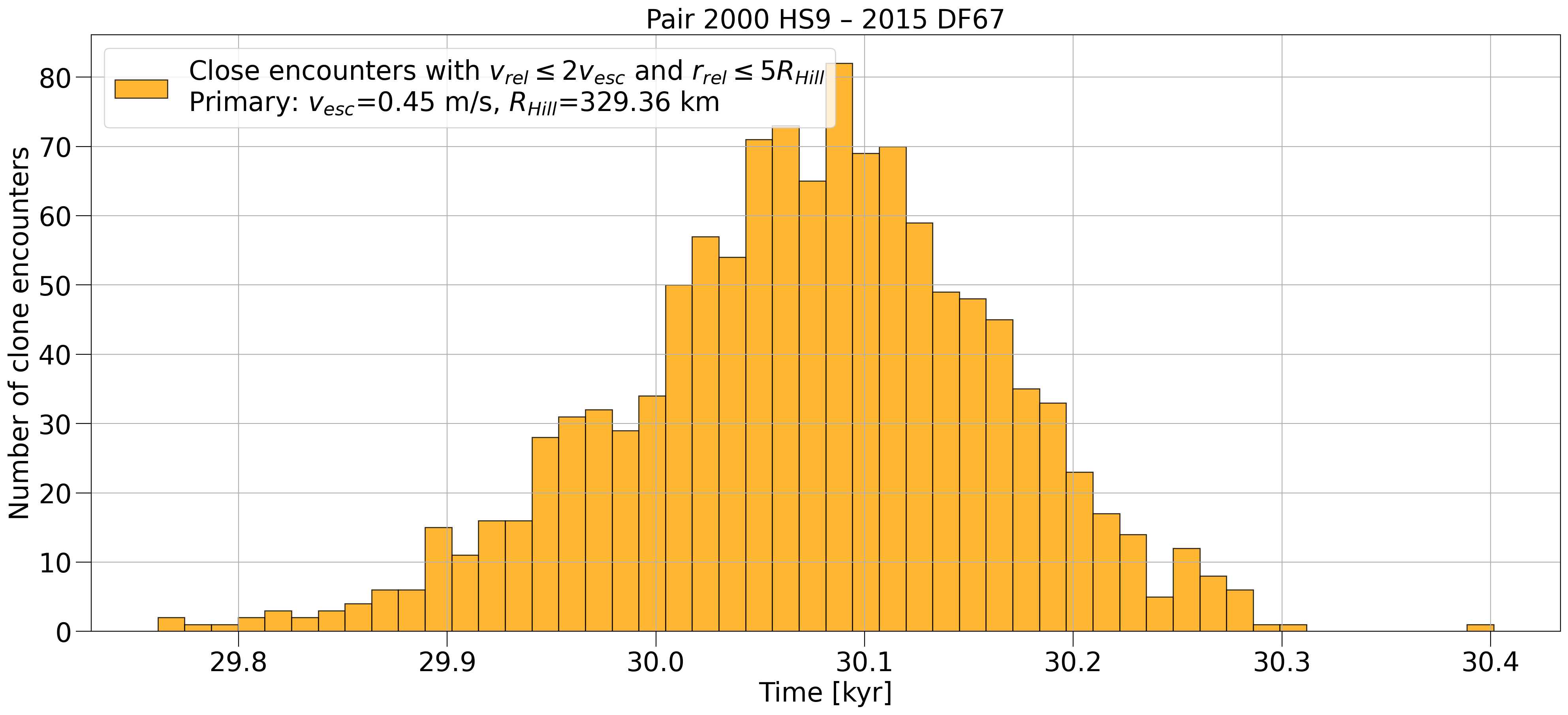}
        \label{fig:figure2}
        \end{minipage}
    \hspace{0.5cm}
        \begin{minipage}[b]{0.49\linewidth} 
        \centering
        \includegraphics[width=\textwidth]{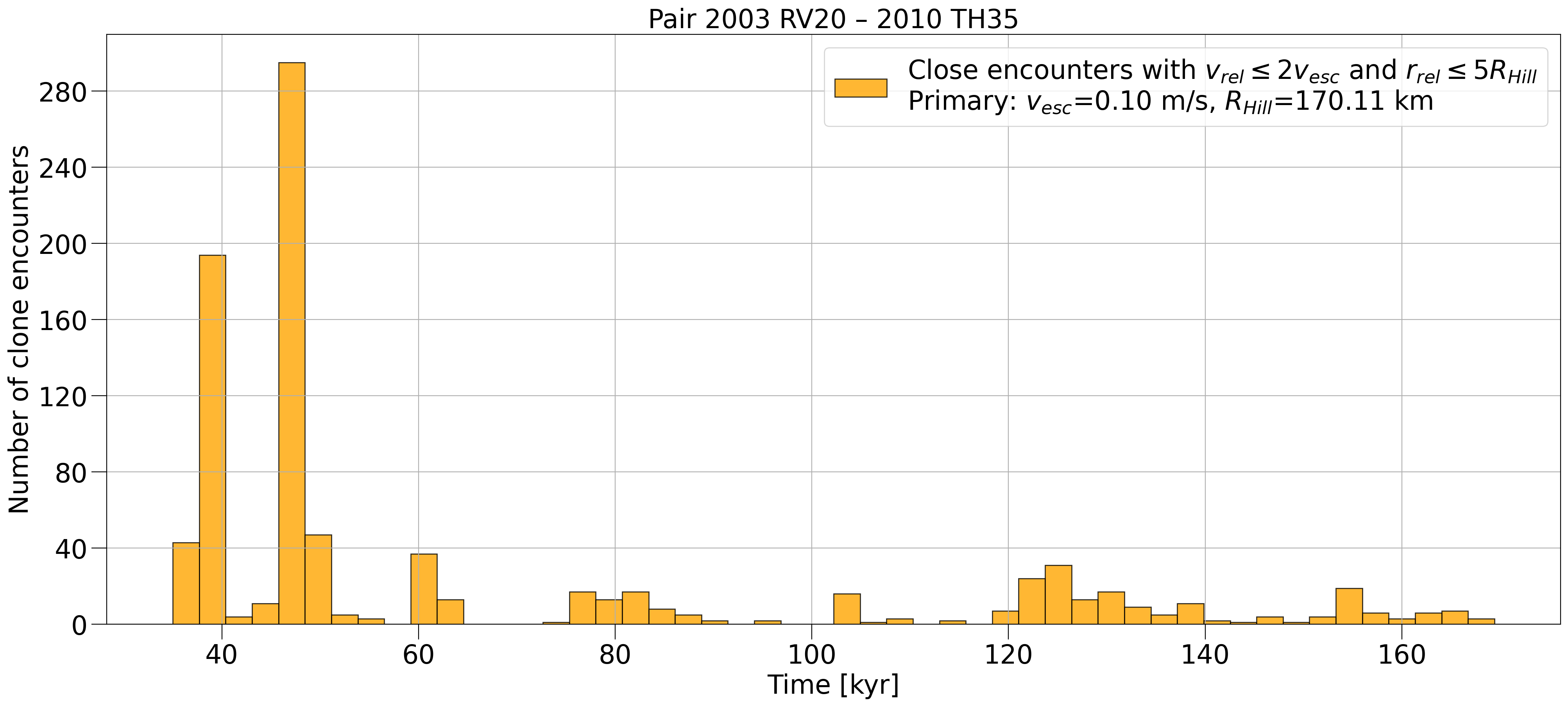}
        \label{fig:figure3}
        \end{minipage}
    \hspace{0.5cm}
     \begin{minipage}[b]{0.49\linewidth} 
        \centering
        \includegraphics[width=\textwidth]{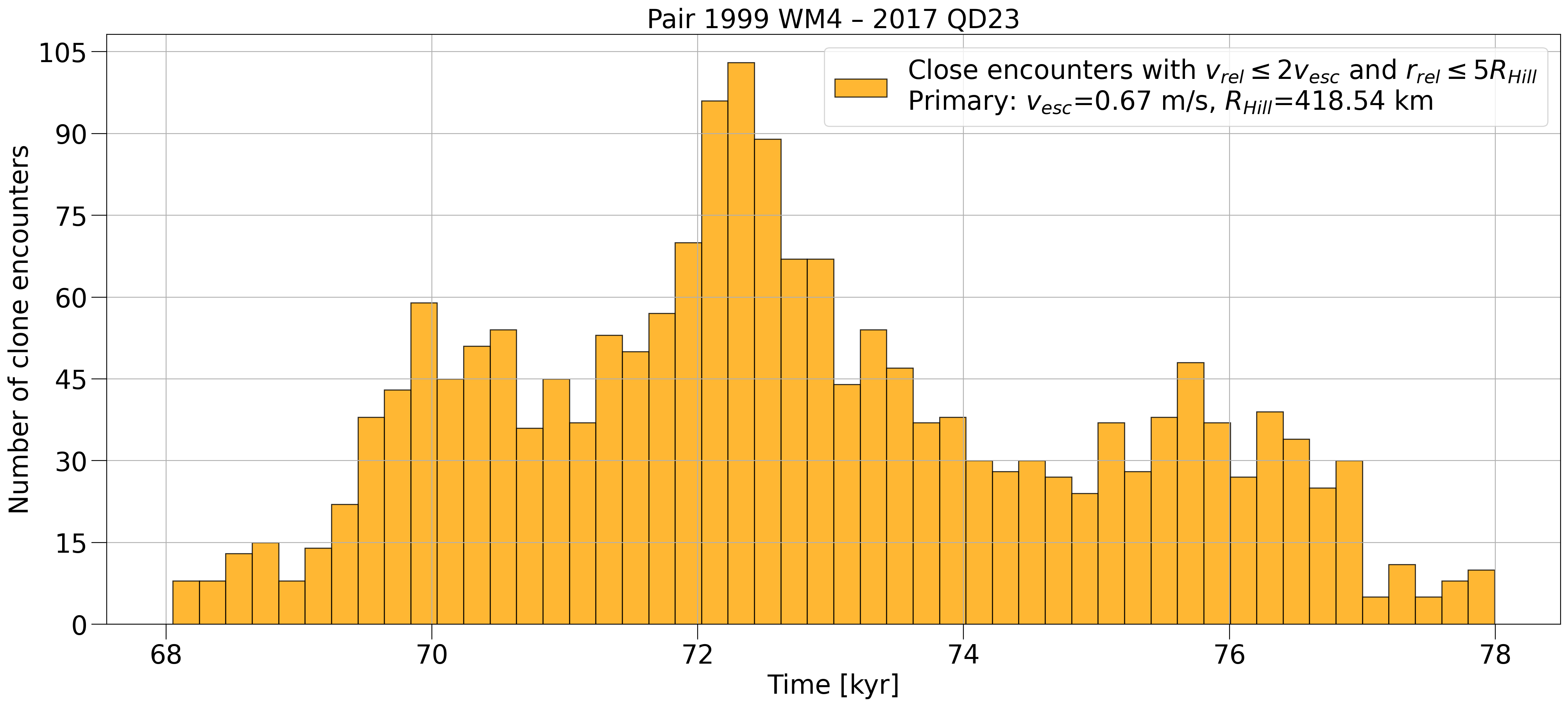}
        \label{fig:figure4}
        \end{minipage}
\caption{\begin{footnotesize} Estimated formation times for the discovered asteroid pair candidates. These distributions include only the clones that had close encounters within specified $v_\mathrm{esc}$ and $R_\mathrm{Hill}$ limits. The limits are shown on the legend of each figure.\end{footnotesize}}
\label{Distribution_times1}
\end{figure*}

\begin{figure*}[ht!]
    \begin{minipage}[b]{0.49\linewidth} 
    \centering
    \includegraphics[width=\textwidth]{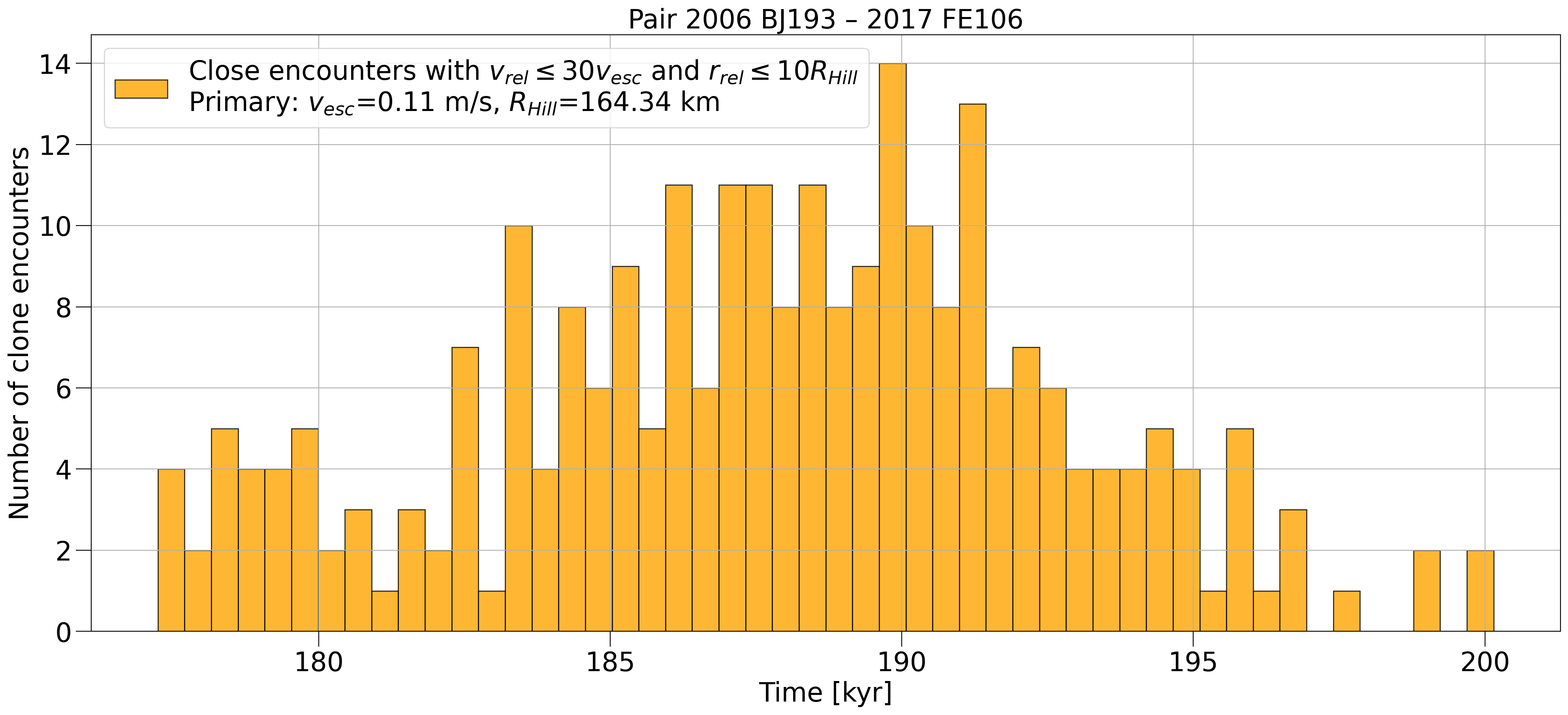}
    \label{fig:figure5}
    \end{minipage}
\hspace{0.5cm}
\begin{minipage}[b]{0.49\linewidth} 
    \centering
    \includegraphics[width=\textwidth]{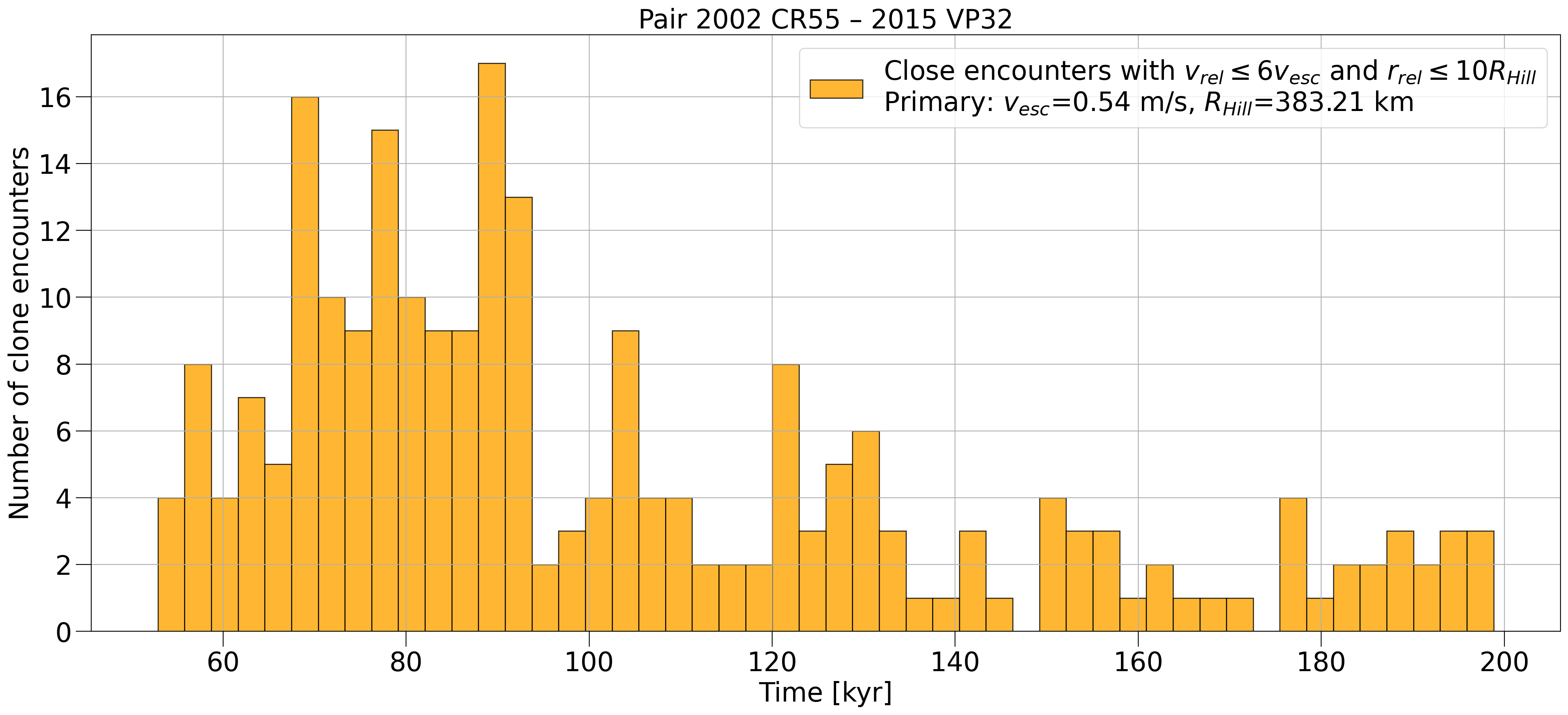}
    \label{fig:figure6}
    \end{minipage}
\hspace{0.5cm}
    \begin{minipage}[b]{0.49\linewidth} 
    \centering
    \includegraphics[width=\textwidth]{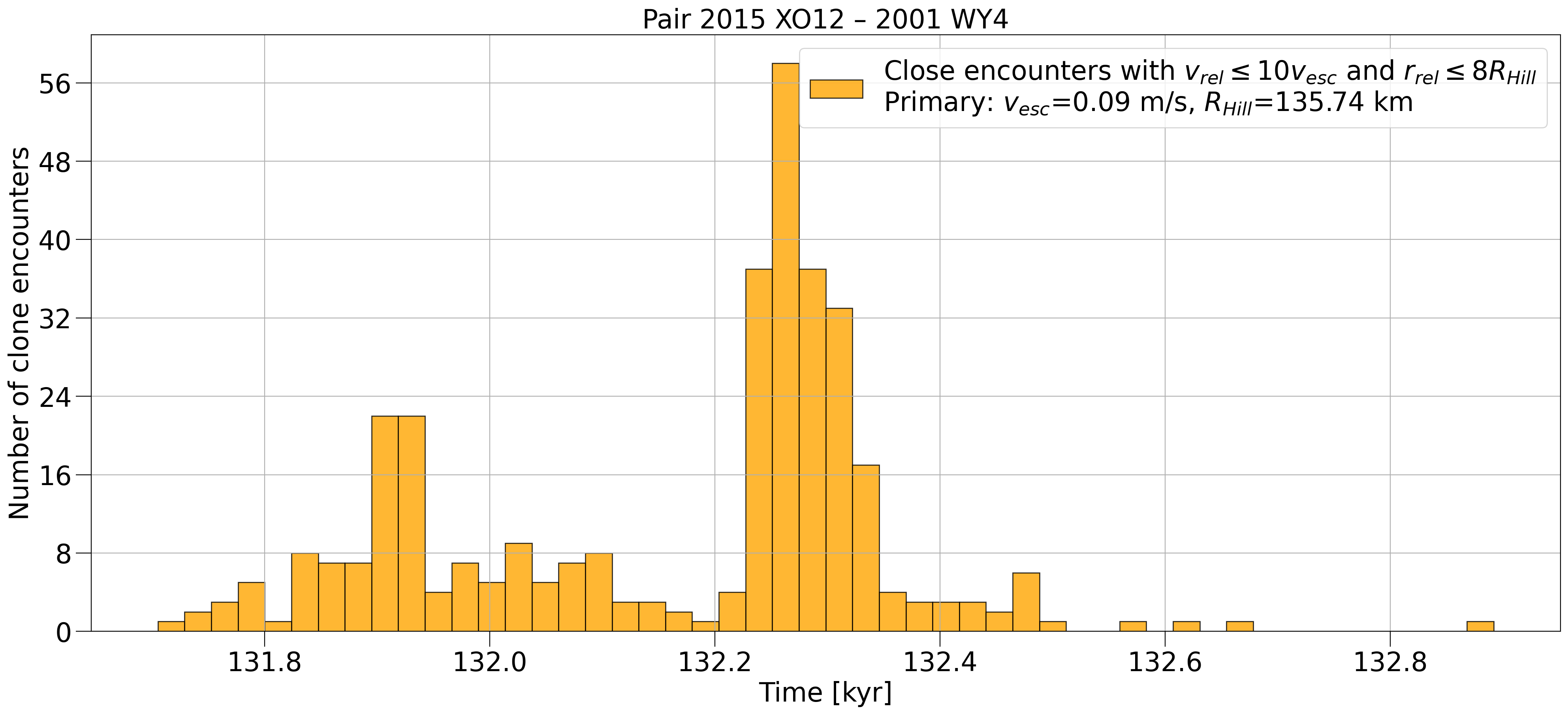}
    \label{fig:figure7}
    \end{minipage}
\hspace{0.5cm}
    \begin{minipage}[b]{0.49\linewidth} 
    \centering
    \includegraphics[width=\textwidth]{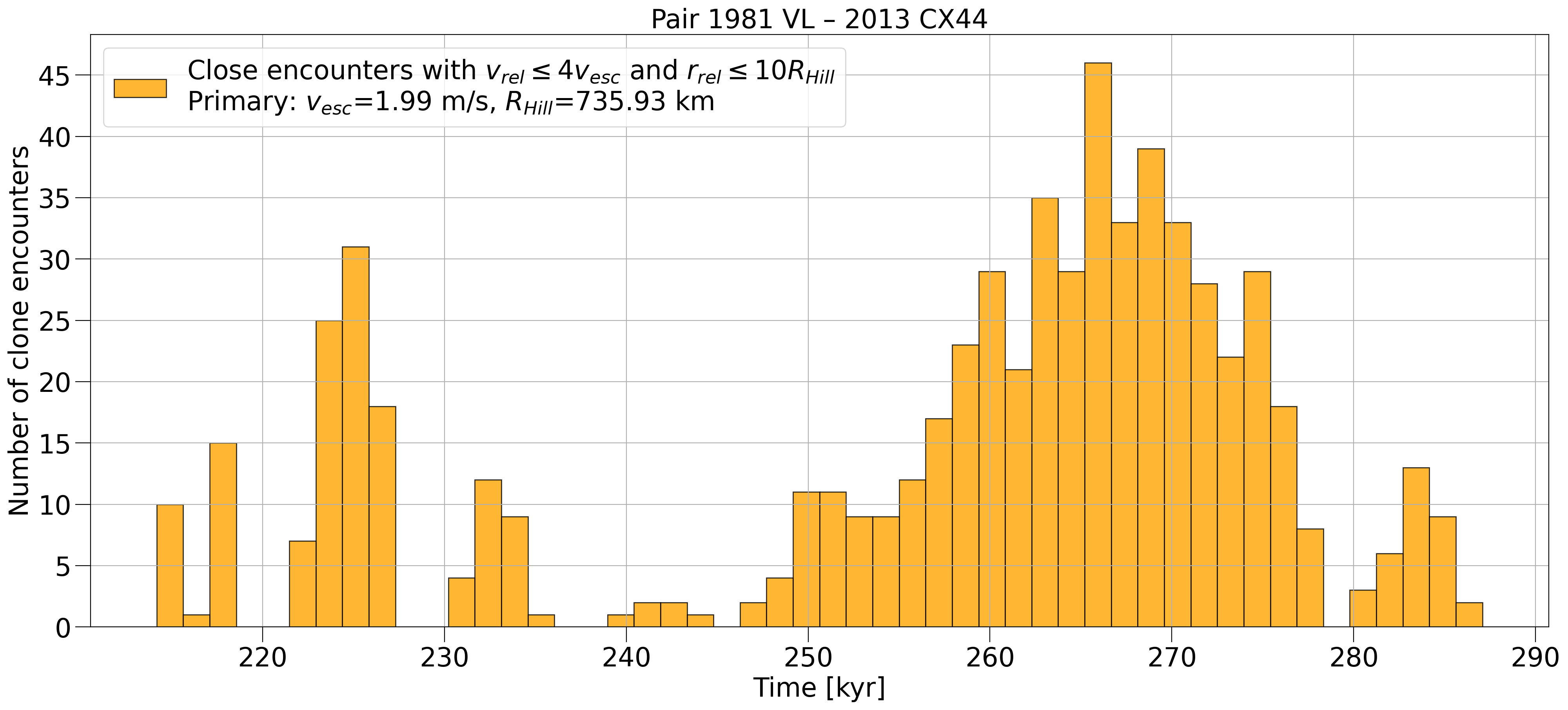}
    \label{fig:figure8}
    \end{minipage}
\hspace{0.5cm}
\begin{minipage}[b]{0.49\linewidth} 
    \centering
    \includegraphics[width=\textwidth]{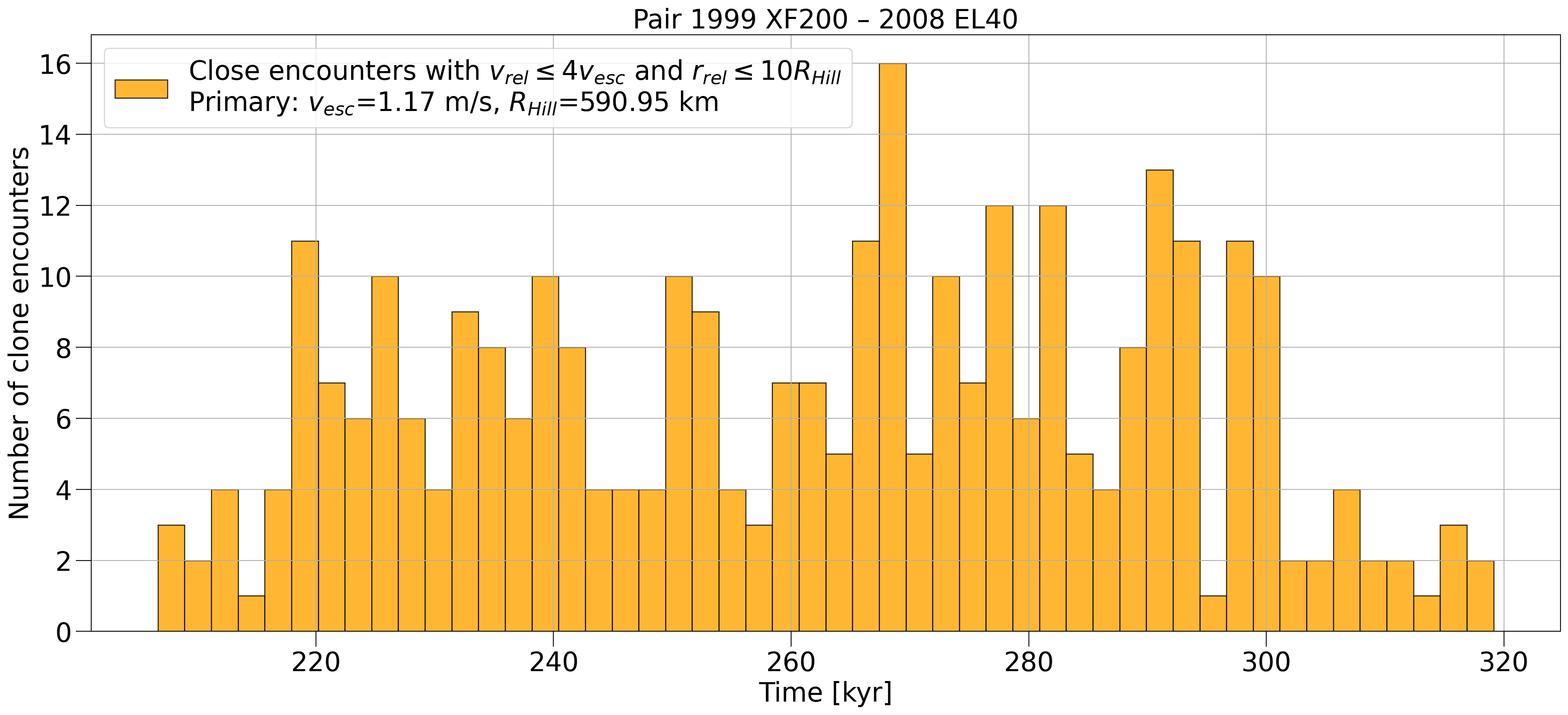}
    \label{fig:figure9}
    \end{minipage}
\hspace{0.5cm}
\begin{minipage}[b]{0.49\linewidth} 
    \centering
    \includegraphics[width=\textwidth]{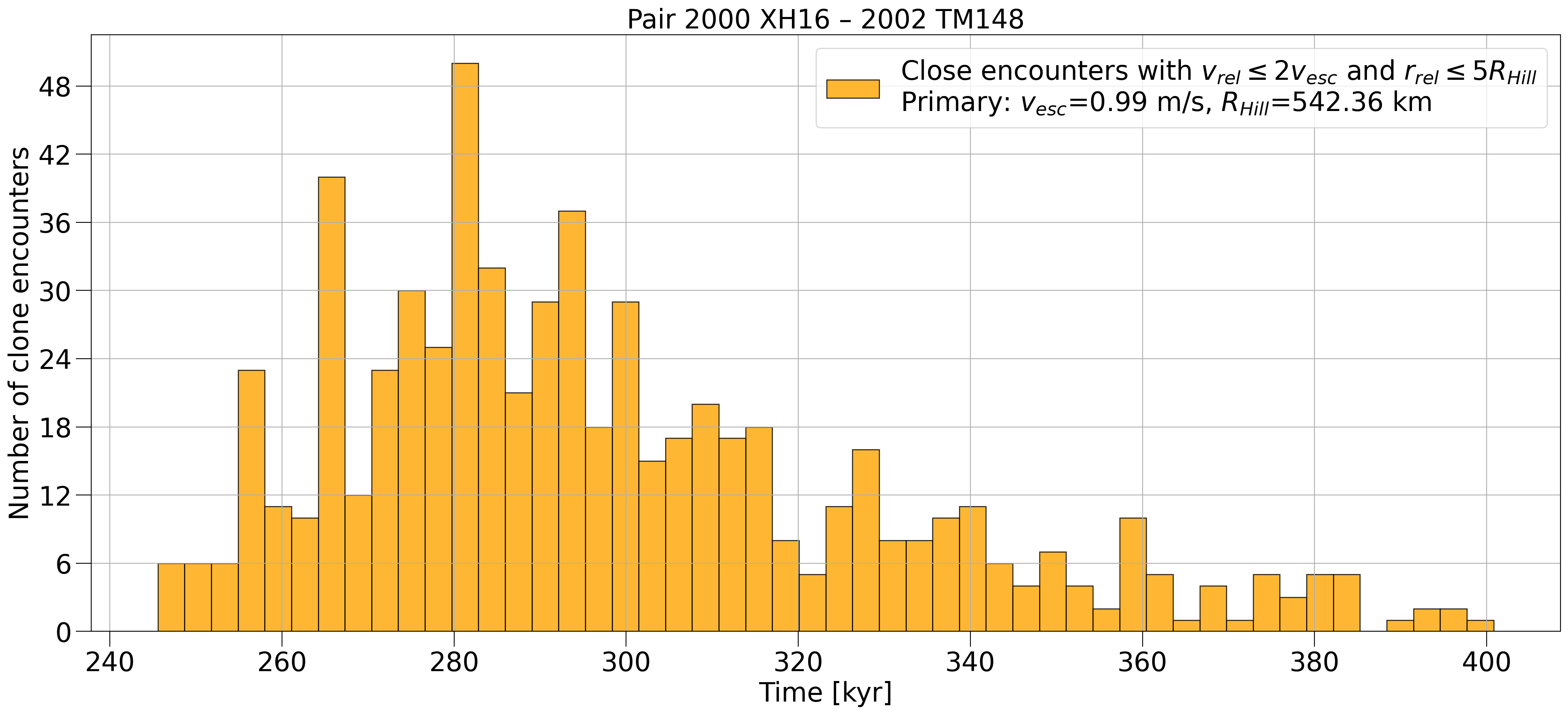}
    \label{fig:figure10}
    \end{minipage}
\hspace{0.5cm}
\caption{\begin{footnotesize} Estimated formation times distribution for the discovered asteroid pair candidates. Continuation of Fig.2.\end{footnotesize}}
\label{Distribution_times2}
\end{figure*}

In the following paragraphs, we describe each found pair candidate consequentially, whereas in Figures \ref{Distribution_times1} and \ref{Distribution_times2} we plot their estimated formation times distribution.
 
 \textbf{1. 271685 (2004 RF90) - 2003 UT336:} \newline
 The asteroid 271685 (2004 RF90) is a primary of the pair with 17.1 mag, and 2003 UT336 is $18.8$ mag, diameter ratio is 1:2.2. 7\% of the clones satisfy the encounter conditions, with multiple encounters for each clone, the closest of which happened 28.1K years ago. Interestingly enough, the distribution has one prominent peak with episodic close encounters following it. The small encounter velocities $v_\mathrm{enc} \sim v_\mathrm{esc}$, and $r_\mathrm{enc} \leq 5r_\mathrm{Hill}$ for this pair indicate that it can be a result of the rotational disruption due to the YORP effect. Using Eqn.(\ref{Spectral classification}), $a_{*}$ for the components was estimated. 2003 UT336 has $a_{*} = 0.18 \pm 0.13$ and 2004 RF90 has $a_{*} = 0.16 \pm 0.07$. While having high uncertainties, this result indicates that the components could be of the same spectral type, and increases the probability of this pair to be a real one. 
 
 \textbf{2. 30243 (2000 HS9) - 2015 DF67:} \newline The components have absolute magnitudes of 16.0 mag for 2000 HS9, and $19.2$ mag for 2015 DF67, with the diameter ratio of 1:4.3. It has a good convergence with 10\% of clones that satisfy close encounter conditions, with multiple encounters. It has one well-determined peak with a small spreading. It is possibly a result of disruption due to the YORP effect. Components 2000 HS9 and 2015 DF67 have the following $a_{*}$ values: $a_{*} =  0.10 \pm 0.04$ and $a_{*} = -0.09 \pm 0.25$ respectively. The value for the secondary component agrees with the primary within its errorbars, but is not constrained enough to make any strong conclusions based on optical properties.
 
 \textbf{3. 405222 (2003 RV20) - 2010 TH35:} \newline 405222 (2003 RV20) is a primary with 17.7 mag, and 2010 TH35 has 18.6 mag, with the diameter ratio of 1:1.5. This pair has a good convergence within $2v_\mathrm{esc}$ and $5R_\mathrm{Hill}$ limits with 10\% of encounters. Most clones are grouped at the age of 48 kyr with multiple smaller peaks of encounters following up to 160 kyr age. Possibly the pair is a result of disruption due to the YORP effect.
 
 \textbf{4.  80245 (1999 WM4) - 540161 (2017 QD23):} \newline The 1999 WM4 is a primary with 15.6 mag, and 2017 QD23 is a secondary component with 18.3 mag, implying the diameter ratio 1:3.5. The resulting encounters are spread around 65K - 75 kyr years region, with 15\% encounters within $2v_\mathrm{esc}$ and $5R_\mathrm{Hill}$ limits. The test for a uniform background asteroid distribution shows that the pair has $P_{1/2} = 0.08$, which is close to the threshold value of 0.05. Therefore, the pair 1999 WM4 - 2017 QD23 is located in a region with some over-density, and potentially may be part of a bigger structure like a small family or a cluster. It also has a high $P_{2}/N_{p}$ ratio, which indicates that it can be a coincidental pair, but results of numerical integration suggest that the members of this pair had close orbital evolution and could be genetically related.
 
 \textbf{5. 204655 (2006 BJ193) - 2017 FE106:} \newline The asteroid 204655 2006 BJ193 is a primary component of this pair with the absolute magnitude of 17.5 mag, while 2017 FE106 is smaller, with only $19.3$ mag. The diameter ratio is 1:2.1. The pair has high median values for the encounter velocity, which is 30 times larger than the estimation for the escape velocity of the primary, and medium relative distances, implying it might be a collisional pair. The 0.9\% of clones satisfy encounter conditions.
 
 \textbf{6. 95251 (2002 CR55) - 2015 VP32:} \newline 2015 VP32 is 18.7 mag, while the primary 2002 CR55 has $H = 15.9$ mag, resulting in the diameter ratio 1:3.6. The clones converge in $1.5\%$ of cases, the encounters are scattered within $10\mathrm{R}_\mathrm{Hill}$, $6v_\mathrm{esc}$ margin.
 
 \textbf{7. 2015 XO12 - 524324 (2001 WY4):}  \newline The primary 2015 XO12 is 17.8 mag, and the secondary 524324 (2001 WY4) has $H=18.4$ mag. The diameter ratio is 1:1.3. The encounters are well grouped in time. Their relative distance is within 8 times the radius of the Hill sphere, and their relative velocity is 10 times the estimated escape velocity of 2015 XO12. 3.5\% of integrations lead to encounters. Possibly the pair is a result of a collisional disruption.
 
 \textbf{8. 13001 Woodney (1981 VL) - 531931 (2013 CX44): \newline} 13001 Woodney (1981 VL) has $H = 14.4$ mag, and the secondary component is 2013 CX44, with a magnitude of 17.9 mag. These magnitudes translate into a 1:5 ratio for the diameters. One of the oldest pairs in our sample, with $5\%$ of clones having multiple encounters scattered along a large time range in the vicinity of $2.5\cdot10^{5}$ yr within $10{R}_\mathrm{Hill}$, $4v_\mathrm{esc}$ margins. 
 
 \textbf{9. 45223 (1999 XF200) - 266505 (2008 EL40):} \newline The 266505 (2008 EL40) has $H = 16.3$ mag, and for 45223 (1999 XF200) additional data are available: $H = 14.7$ mag, diameter = $2.9 \pm 0.2$ km, rot. period = 4.903 hr, albedo = $0.30 \pm 0.04$). The diameter ratio for the pair is 1:1.8, 3\% of encounters is within $10\mathrm{R}_\mathrm{Hill}$, $4v_\mathrm{esc}$ margin. Possibly this pair is a result of disruption due to the YORP effect.
 
 \textbf{10. 67982 (2000 XH16) - 317521 (2002 TM148):} \newline The primary 67982 (2000 XH16) has the following parameters: $H = 15.0$ mag, diameter = $2.7 \pm 0.3$ km, albedo = $0.33 \pm 0.05$, and the secondary 317521 (2002 TM148) is 16.7 mag. The components diameter ratio is 1:2.6\% of encounters is within $5{R}_\mathrm{Hill}$, $2v_\mathrm{esc}$ margin. Possibly this pair is a result of disruption due to the YORP effect.

 \begin{table*}[h!]
 \captionsetup{justification=centering}
 \caption{Main data on new pair candidates. For the formation age estimates uncertainties are calculated as 5th and 95th percentiles of the formation age distribution. Acceptable values for $P_{1/2}$ are considered to be 0.01 and larger for pairs with $d$ < 10 m/s, and 0.05 for pairs with greater $d$. For the ratio $P_{2}/N_{p}$ some contamination by coincidental pairs is expected with $P_{2}/N_{p} = 0.1$ and larger.}
    \setlength{\tabcolsep}{15pt}
    \renewcommand{\arraystretch}{1.3}
    \centering
        \begin{tabular}{|p{0.2cm}|p{1.9cm}|p{1.75cm}|p{1.0cm}|p{0.8cm}|p{0.7cm}|p{1.0cm}|p{0.6cm}|p{0.9cm}|}
         \hline
         \multicolumn{9}{|c|}{Properties of asteroid candidates} \\
         \hline
         No. & asteroid name & color index $a_{*}$ & absolute magnitude H & diameter ratio & $d_{mean}$ (m/s) & est.age (kyr) & $P_{1/2}$  & $P_{2}/N_{p}$ \\
         \hline
         1 & 2004 RF90  & $0.16\pm0.07$& 17.1 & 1:2.2 & 13.5 & $30.1_{-2.0}^{+4.6}$ & 1 & 0.068\\
         - & 2003 UT336 & $0.19\pm0.13$& 18.8 &  - &&&& \\
         
         2 & 2000 HS9 & $0.10\pm0.04$ & 16.0 & 1:4.3 & 22.6 & $30.1_{-0.15}^{+0.15}$ & 0.663 & 0.0003\\
         - & 2015 DF67 & $-0.09\pm0.25$ & 19.2 &&&&& \\
         
         3 & 2003 RV20 & - & 17.7 & 1:1.5 & 9.39 & $46.0^{+106.6}_{-8.3}$ & 0.012 & 0.047\\
         - & 2010 TH35 & - & 18.6 &&&&& \\
         
         4 & 1999 WM4 & - & 15.6 & 1:3.5 & 9.93 & $72.4^{+4.1}_{-2.9}$ & 0.081 & 0.578\\  
         - & 2017 QD23 & - & 18.3 &&&&& \\
         
         5 & 2002 CR55 & - & 15.9 & 1:3.6 &6.94 & $90.1^{+94.3}_{-31.3}$ & 0.423 & 0.002\\
         - & 2015 VP32 & - & 18.7 &&&&& \\
         
         6 & 2015 XO12 & $-0.03\pm0.05$ & 17.8 & 1:1.3 & 1.86 & $132.2^{+0.2}_{-0.4}$ & 1 & 8e-05\\ 
         - & 2001 WY4 & - & 18.4 &&&&&\\
         
         7 & 2006 BJ193 & $0.06\pm1.08$ & 17.5 & 1:2.1 & 23.5 & $188^{+7.5}_{-9.3}$ & 0.154 &  0.071\\  
         - & 2017 FE106 & - & 19.3 &&&&& \\
         
         8 & 1981 VL & - & 14.4 & 1:5 & 22.2 & $263.7^{+16.3}_{-40.8}$ & 0.395 & 0.123\\
         - & 2013 CX44 & - & 17.9 &&&&& \\
         
         9 & 1999 XF200 & - & 14.7 & 1:1.8 & 14.9 & $265.8^{+37.7}_{-47.5}$ &  0.345 & 0.083\\ 
         - & 2008 EL40 & $0.20\pm0.04$&16.3 &&&&& \\
         
         10 & 2000 XH16 & $0.14\pm0.03$ & 15.0 & 1:2 & 14.6 & $291.8^{+70.1}_{-35.3}$ & 0.393 & 0.028\\ 
         - & 2002 TM148 & - & 16.7 &&&&& \\
         \hline
         \end{tabular}
 \label{New_pairs_age}
 \end{table*}
 
\section{Conclusions}
    As a result of this work, we have developed a new pipeline for asteroid pairs search. The working capacity of the pipeline was proved on a sample of known asteroid pairs, with the good agreement of our calculations and the results of the previous studies. A survey of the inner part of the main belt was performed, revealing 10 new asteroid pairs. Pair candidates were initially identified in a five-dimensional space of osculating orbital elements. Then the candidates were verified using numerical modeling that took into account a gravitational influence, as well as the non-gravitational Yarkovsky effect in its simplified form. 
    
    The obtained pairs were checked for uniform background asteroid distribution and their statistical significance. All the pairs passed the test for uniform background asteroid distribution, with pairs 2003 RV20 - 2010 TH35 and 1999 WM4 - 2017 QD23 having $P_{1/2}$ close to the threshold values. This may indicate that they are located in an overdensity region, and each of them may be a part of a bigger structure like a family or cluster. The 1999 WM4 - 2017 QD23 pair also has high $P_{2}/N_{p}$ ratio, indicating that this pair can be coincidental, but results of numerical integration suggest that the members of this pair had close orbital evolution and are possibly genetically related.
    
    For all the pairs, possible formation ages were found to be within the 400 kyr limit, with typical errors in age estimation of 10-20 percent. This bias for younger pairs could be due to the used technique of candidate selection, which originates from the small $d$ on the pre-selection stage and gets further enhanced by the imposed requirement of good convergence of the preliminarily tested clones. The absence of pairs discovered by our pipeline beyond 400 kyr came as a surprise for us, but the successful recovery of known pairs over this limit persuades us that the pipeline is up to recovery of older pairs as well.
    
    Most pairs in our sample contain at least one asteroid discovered during the last decade. This can explain why these pairs were not found in previous studies. Typical values of $v_\mathrm{esc}$ and $R_\mathrm{Hill}$ for the discovered pair candidates in the case of close encounters hint the influence of the YORP effect to be the main catalyst of their formation.
    
    The color indices data were obtained for two pairs, giving some insight into spectral class similarity for the pairs' components. Pair 2003 UT336 - 271685 (2004 RF90) has matching color indices $a_{*}$, which suggests the possibility of the same spectral class for its components. The color indices for 30243 (2000 HS9) - 2015 DF67 are matching within errors, although the errors are large and unconstraining. 
    
    The first results of our pipeline presented in this paper are the initial steps of our survey of asteroid pairs. A number of major advancements are planned for the future, most important of which is the improved simulation of the Yarkovsky force motivated by thermal models of individual asteroids. The enhanced accuracy and reliability of our survey will allow us to focus on the origin of asteroid pairs.
    Their key formation mechanisms, namely the collisional disruption and rotational fission, can differently manifest themselves in the convergence of the orbits of the pair members, with a closer approach in coordinates or velocities correspondingly. Thus the extensive statistics on discovered pairs and the more precise simulation of their dynamics will allow us to shed light on their genesis. Moreover, in the case of the unknown Yarkovsky effect, the mere assumption of the pair's convergence in the past can serve to constrain the value of the non-gravitational acceleration, thus providing a new tool for measuring the Yarkovsky effect.
    
\begin{acknowledgements}
The authors thank Alexey Sergeyev and Ivan Slyusarev for sharing their results on the colors of the asteroids from our sample, to Petr Pravec for clarification of the pair significance analysis, and to the reviewer of an article Petr Fatka for multiple insightful suggestions that lead to a significant improvement of the article. 
The publication has been prepared as a result of the project N2020.02/0371 “Metallic asteroids: search for parent bodies of iron meteorites, sources of extraterrestrial resources” financed by the National Research Foundation of Ukraine using the state budget.
\end{acknowledgements}

\appendix
\section{Estimate of the Yarkovsky effect}
\label{AppendixYarkovsky}
To estimate the Yarkovsky effect experienced by an asteroid, we use the theory proposed in \cite{golubov16}. We apply Eqn. (41) from \cite{golubov16} to a spherical body, resulting in the following expression for the Yarkovsky force,
\begin{equation}
F_\mathrm{Yark}=\frac{(1-A)\Phi R^2}{c}\int\limits_\frac{\pi}{2}^{-\frac{\pi}{2}}2\pi\cos{\Psi}p_\mathrm{Yark}^\tau\,d\Psi.
\label{FYark16}
\end{equation}

The non-dimensional Yarkovsky pressure $p_\mathrm{Yark}^\tau$ is determined numerically, and the results are plotted in Figure 4 in \cite{golubov16}. By fitting green lines in the three right-hand panels of this figure, one gets an approximate expression for $p_\mathrm{Yark}^\tau$ at $\theta=1$,
\begin{equation}
p_\mathrm{Yark}^\tau\approx0.033\cos\varepsilon\cos^2\Psi.
\label{pYark}
\end{equation}
For the thermal parameters $\theta$ that are far away from unity, $p_\mathrm{Yark}$ can be substantially less. Still, from
Table 1 in \cite{golubov12} we can see that for the main belt asteroids covered with regolith we have $\theta\sim 1$, thus $p_\mathrm{Yark}^\tau$ is close to or somewhat less than Eqn. (\ref{pYark}). 

Substituting Eqn. (\ref{pYark}) into Eqn. (\ref{FYark16}), performing the integration, and assuming $\varepsilon=0$, we get
\begin{equation}
F_\mathrm{Yark0}\approx 0.033\frac{8\pi}{3}\frac{(1-A)\Phi R^2}{c}=0.28\frac{(1-A)\Phi R^2}{c}.
\end{equation}
If $\varepsilon$ is large or $\theta$ is far away from unity, the actual value of the Yarkovsky effect can be anywhere in the range between $-F_\mathrm{Yark0}$ and $F_\mathrm{Yark0}$.
\end{document}